\documentclass[
reprint,
dcolumn,
bm,
nofootinbib,
aps,
pra,
notitlepage]{revtex4-1}
\renewcommand{\baselinestretch}{1.0}
\usepackage{epsfig}
\usepackage{color}

\usepackage{graphicx}
\usepackage{color}
\usepackage[latin1]{inputenc}
\usepackage[T1]{fontenc}
\usepackage{amsmath}
\usepackage{amssymb}
\usepackage{braket}
\usepackage{mathrsfs}
\usepackage{hyperref}
\usepackage{upgreek}
\usepackage{pstricks-add}
\usepackage{natbib} 
\usepackage{listings}
\usepackage{tikz}
\usetikzlibrary{mindmap}
\begin{document}
\title{Cavity-Catalyzed Hydrogen Transfer Dynamics in an Entangled Molecular Ensemble under Vibrational Strong Coupling}
\author{Eric W. Fischer}
\email{ericwfischer.sci@posteo.de}
\affiliation{Theoretische Chemie, Institut f\"ur Chemie, Universit\"at Potsdam,
Karl-Liebknecht-Strasse 24-25, D-14476 Potsdam-Golm, Germany}

\author{Peter Saalfrank}
\affiliation{Theoretische Chemie, Institut f\"ur Chemie, Universit\"at Potsdam,
Karl-Liebknecht-Strasse 24-25, D-14476 Potsdam-Golm, Germany}
\affiliation{Institut f\"ur Physik und Astronomie, Universit\"at Potsdam, Karl-Liebknecht-Stra\ss e 24-25, D-14476 Potsdam-Golm, Germany}

\let\newpage\relax

\begin{abstract}
Microcavities have been shown to influence the reactivity of molecular ensembles by strong coupling of molecular vibrations to quantized cavity modes. In quantum mechanical treatments of such scenarios, frequently idealized models with single molecules and scaled, effective molecule-cavity interactions or alternatively ensemble models with simplified model Hamiltonians are used. In this work, we go beyond these models by applying an ensemble variant of the Pauli-Fierz Hamiltonian for vibro-polaritonic chemistry and numerically solve the underlying time-dependent Schr\"odinger equation to study the cavity-induced quantum dynamics in an ensemble of thioacetylacetone (TAA) molecules undergoing hydrogen transfer under vibrational strong coupling (VSC) conditions. Beginning with a single molecule coupled to a single cavity mode, we show that the cavity indeed enforces hydrogen transfer from an enol to an enethiol configuration with transfer rates significantly increasing with light-matter interaction strength. This positive effect of the cavity on reaction rates is different from several other systems studied so far, where a retarding effect of the cavity on rates was found. It is argued that the cavity ``catalyzes'' the reaction by transfer of virtual photons to the molecule. The same concept applies to ensembles with up to $N=20$ TAA molecules coupled to a single cavity mode, where an additional, significant, ensemble-induced collective isomerization rate enhancement is found. The latter is traced back to complex entanglement dynamics of the ensemble, which we quantify by means of von Neumann-entropies. A non-trivial dependence of the dynamics on ensemble size is found, clearly beyond scaled single-molecule models, which we interpret as transition from a multi-mode Rabi to a system-bath-type regime as $N$ increases. 
\end{abstract}

\let\newpage\relax
\maketitle
\newpage
\section{Introduction}
The interaction of optical modes confined in Fabry-P\'erot cavities with optically active molecular degrees of freedom lies at the heart of the emerging field of polaritonic chemistry\cite{ebbesen2016,ribeiro2018,feist2018,dunkelberger2022,yuenzhou2022,fregoni2022,li2022}. In one class of promising experiments, molecular vibrations interact strongly with the ground state of infrared active cavity modes. In this \textit{vibrational strong coupling} (VSC) regime, significantly altered thermal ground state chemistry has been observed\cite{george2015,thomas2016,thomas2019,lather2019,thomas2020}. While in early examples of VSC-altered reactivity, rates usually were retarded, meanwhile also experiments exist with accelerated rates\cite{ebbesen2021}. VSC experiments have generated a significant theoretical effort aiming to understand the complex interplay of light and matter in thermal polariton chemistry\cite{galego2019,campos2019,campos2020,li2020,vurgaftman2020,wiesehan2021,imperatore2021,li2021,yang2021,sun2022,lindoy2022,fischer2022,mandal2022,wellnitz2022,mandal2022b}. As an example, isomerization reactions have been idealized as arising from population transfer from one well of a cavity-distorted double-minimum potential to the other well, with the dynamics being treated classically or quantum mechanically\cite{li2021,sun2022,lindoy2022,fischer2022}. At least in quantum mechanical treatments, typically single molecules are considered and the transition to ensembles of $N$ molecules is mimicked \textit{via} effective single-molecule models with scaled molecule-cavity couplings\cite{fischer2022}. Alternatively, simplified $N$-emitter model Hamiltonians comprising a set of two-level systems and model cavity-emitter couplings are used such as the Tavis-Cummings\cite{tavis1968} or Dicke\cite{dicke1954}-models.

In this contribution, we study hydrogen transfer reaction models for thioacetylacetone (TAA) molecules\cite{doslic1999} placed in an infrared cavity both for a single molecule or an ensemble of up to $N=20$ molecules. The latter are explicitly treated from a fully quantum mechanical perspective and beyond simplified $N$-emitter model Hamiltonians such as Tavis-Cummings\cite{tavis1968}, by employing an $N$-molecule variant of the Pauli-Fierz Hamiltonian and solving a corresponding multi-dimensional time-dependent Schr\"odinger equation numerically.
For the molecule under investigation, idealized here as an asymmetric double-well, we first demonstrate for the single-molecule scenario a cavity-induced isomerization/\,H-transfer from the ``enol'' configuration, which is the energetically favored form outside the cavity, to an ``enethiol'' form. We extract approximate H-transfer rates from the numerically obtained dynamics, which increase with light-matter coupling and ensemble size. The cavity ``catalytic'' effect in the single-molecule limit, \textit{i.e.}, rate enhancement, is traced back to a \textit{symmetric} distortion/\,displacement of the cavity potential energy surface (cPES) by a smoothly varying dipole function. In contrast, if the distortion of the cPES is \textit{antisymmetric} (as for the inversion of ammonia, for example, with an antisymmetric dipole function), rates are found to be decelerated by the cavity\cite{fischer2022}. Equivalently, the rate enhancement found here can be understood as being driven by transfer of virtual photons from the cavity to the H-transfer system. The concept of virtual photon transfer directly generalizes to the ensemble scenario, where single-molecule cPES distortion arguments no longer hold due to significantly reduced \textit{single} molecule light-matter coupling in contrast to enhanced ensemble coupling.\cite{sun2022} In particular, we discuss how virtual photon transfer in the many-molecule scenario leads to a strongly entangled molecular transfer ensemble, which in turn determines the quantum mechanical nature of the transfer dynamics, beyond those arising from scaled single-molecule model Hamiltonians.

The paper is organized as follows. In Sec.\ref{sec2} the $N$-molecule Pauli-Fierz Hamiltonian for an ensemble of TAA molecules is introduced and the computation of observables employed to describe the isomerization dynamics is illustrated. In Sec.\ref{sec3}, we demonstrate and analyze the cavity-induced isomerization of single TAA and ensembles of TAA molecules systematically in dependence of the molecule-cavity coupling strength and the ensemble size. Finally, Sec.\ref{sec4} summarizes this work. Most of the numerical details and parameters as well as some further results can be found in the Supplementary Information (SI).

\section{Theory and Model}
\label{sec2}
\subsection{Hamiltonian and Quantum Dynamics}
\begin{figure*}[hbt!]
\includegraphics[scale=1.0]{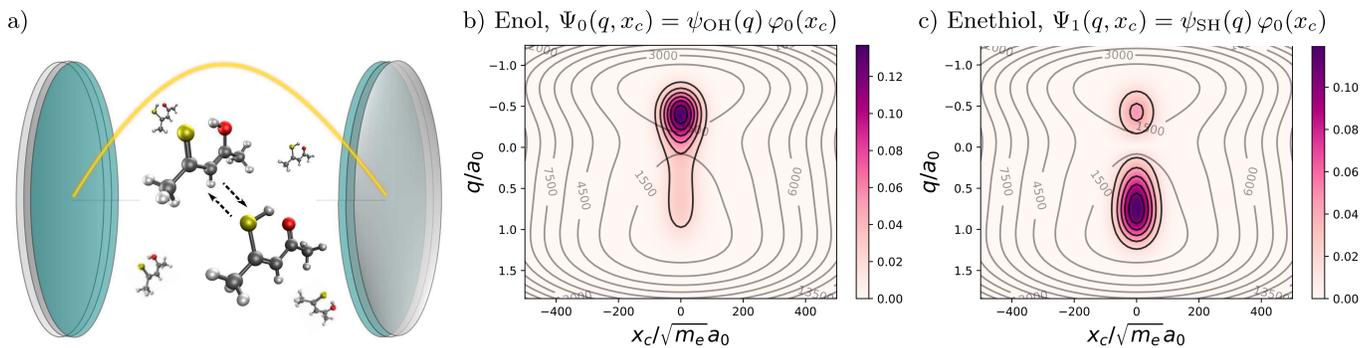}
\renewcommand{\baselinestretch}{1.}
\caption{(a) Schematic setup for thioacetylacetone (TAA) isomerization in a single-mode cavity model. Indicated are the enol (with OH group, O in red) and enethiol configurations (with SH group, S in yellow), other molecules in the ensemble, and the cavity mode between two parallel mirrors. b) Ground state density, $\vert\Psi_0(q,x_c)\vert^2$, corresponding mainly to the enol configuration and c) first excited state density, $\vert\Psi_1(q,x_c)\vert^2$, corresponding mainly to the enethiol configuration, both embedded in a two-dimensional cPES, $V(q,x_c)$ (contours in $\mathrm{cm}^{-1}$), which arises from a cavity-hydrogen-transfer model Hamiltonian. The single-molecule cPES is a function of the H-transfer coordinate, $q$, and a cavity displacement coordinate, $x_c$. The uncoupled case with $\eta=0.0$ is shown with lowest-state energies, $E_0=\varepsilon_0 +\frac{\hbar\omega_c}{2}$, and, $E_1=\varepsilon_1+\frac{\hbar\omega_c}{2}$, where $\varepsilon_0=966.3\,\mathrm{cm}^{-1}$ and $\varepsilon_1=1092.8\,\mathrm{cm}^{-1}$ are the lowest two eigenenergies of the one-dimensional H-transfer Hamiltonian, $\hat{H}_S$, with potential, $V(q)$. We set the cavity frequency to be resonant with the lowest molecular vibrational transition, \textit{i.e.}, $\hbar\omega_c=\Delta \varepsilon_{10}=126.5\,\mathrm{cm}^{-1}$.} 
\label{fig.htransfer_densities}
\end{figure*}

We consider a cavity-altered asymmetric hydrogen transfer reaction by extending a well studied reaction model Hamiltonian for TAA\cite{doslic1999} (\textit{cf.} Fig.\ref{fig.htransfer_densities}a). The light-matter hybrid system is described by an effective Pauli-Fierz Hamiltonian in length-gauge representation, cavity Born-Oppenheimer type and long-wavelength approximations, which reads\cite{flick2017a,schaefer2018,fischer2021}
\begin{equation}
\hat{H}
=
\hat{H}_S
+
\hat{H}_C
+
\hat{H}_{SC}
+
\hat{H}_{DSE}
\quad.
\label{htot}
\end{equation}
The first term resembles $N$ non-interacting H-transfer systems, idealized here as a sum of one-dimensional Hamiltonians along a transfer coordinate, $q_i$, for the $i$-th molecule with 
\begin{equation}
\hat{H}_S
=
\sum^N_{i=1}
\biggl(
-
\dfrac{\hbar^2}{2\mu_S}\dfrac{\partial^2}{\partial q^2_i}
+
V(q_i)
\biggr)
\quad,
\label{eq.ensemble_transfer_hamiltonian}
\end{equation}
and corresponding reduced mass, $\mu_S$, close to the hydrogen mass (\textit{cf.} SI, Sec.I). As shown in the SI, Sec.I, this Hamiltonian is a one-dimensional approximation obtained from a two-dimensional Hamiltonian developed originally in Ref.\cite{doslic1999}. The single-molecule potential, $V(q)$ (where we suppress the index $i$ here, as all potentials are identical), constitutes an asymmetric double-well potential with a global minimum at $q=-0.572\,a_0$, which relates to the enol (OH) configuration, and a local minimum at $q=0.947\,a_0$ corresponding to the enethiol (SH) configuration of TAA. Both minima are characterized by classical over-the-barrier activation energies, $\Delta E^{cl}_\mathrm{OH}=1598\,\mathrm{cm}^{-1}$ and $\Delta E^{cl}_\mathrm{SH}=1081\,\mathrm{cm}^{-1}$, with respect to the transition state located at $q=0.0$, \textit{i.e.}, the classical energy difference between the two isomers is $517\,\mathrm{cm}^{-1}$. By diagonalizing $\hat{H}_S$ for the single-molecule case with $N=1$, the two energetically lowest lying eigenstates are found to correspond to the ground state enol configuration, $\psi_0(q)=\psi_\mathrm{OH}(q)$, and the first excited state, enethiol configuration, $\psi_1(q)=\psi_\mathrm{SH}(q)$, with energies, $\varepsilon_0=966.3\,\mathrm{cm}^{-1}$ and $\varepsilon_0=1092.8\,\mathrm{cm}^{-1}$, respectively, giving a corresponding quantum mechanical energy difference of $\Delta \varepsilon_{10}=126.5\,\mathrm{cm}^{-1}$. Details on the transfer potential, $V(q)$, with all numerical parameters are provided in the SI, Sec.I. 

The second term in Eq.(\ref{htot}) describes the cavity Hamiltonian, which we restrict here to a single mode, given in coordinate representation as
\begin{equation}
\hat{H}_C
=
-
\dfrac{\hbar^2}{2}
\dfrac{\partial^2}{\partial x^2_c}
+
\dfrac{\omega^2_c}{2}
x^2_c
\quad,
\label{eq.cavity_hamiltonian}
\end{equation}
with cavity displacement coordinate, $x_c$ (of dimension mass$^{1/2} \times$ length), and harmonic cavity frequency, $\omega_c$, chosen to be resonant with the lowest vibrational transition of the transfer system, \textit{i.e.}, $\hbar\omega_c=\Delta \varepsilon_{10}=126.5\,\mathrm{cm}^{-1}$. Further, the light-matter interaction term, $\hat{H}_{SC}$, is given by 
\begin{equation}
\hat{H}_{SC}
=
\sqrt{\dfrac{2\omega_c}{\hbar}}\,
g_N\,
d(\underline{q})\,
x_c
\quad,
\label{eq.light_matter_interaction}
\end{equation}
with collective transfer coordinate, $\underline{q}=(q_1\dots q_N)$. While going well beyond Dicke-Hamiltonians, the collective light-matter interaction constant, $g_N$, is still chosen to be of Dicke form\cite{dicke1954}, $g_N=\frac{g}{\sqrt{N}}$, where $g$ is a single-molecule coupling strength. Note, we do not take into account any cavity size effects, since we work in the long-wavelength approximation, and treat $g$ in the spirit of a variable effective model coupling parameter not accounting for its dependence on the cavity volume, $g\propto\frac{1}{\sqrt{V_c}}$. We also do not account for solvent effects, which are expected to mainly open energy transfer channels, at least on the timescale of picoseconds.

The collective dipole moment is given by, $d(\underline{q})=\sum^N_i d(q_i)$, with ground state dipole function, $d(q_i)$, for the $i$-th H-transfer system. The latter is a linear approximation to the dipole function given in Ref.\cite{doslic1999} and reads
\begin{equation}
d(q_i)
=
d_0
+
d_S\,
(q_i-q_0)
\quad,
\end{equation}
with parameters $d_0$, $d_S$ and $q_0$ given in the SI, Sec. I. We note that the dipole moment takes positive values at both the enol minimum ($d_\mathrm{OH}=1.678\,ea_0$) and the enethiol minimum ($d_\mathrm{SH}=1.482\,ea_0$), \textit{i.e.}, $d(q_i)$ changes smoothly without sign change along the (classical) reaction path. Note further that in Eq.(\ref{eq.light_matter_interaction}), we assume the dipole moment of each molecule to be aligned with both the respective H-transfer coordinate and the polarization axis of the cavity mode. The single-molecule coupling constant, $g$, in Eq.\eqref{eq.light_matter_interaction} has dimension of an electric field strength and is modeled here as, $g=\frac{\hbar\omega_c}{d_{10}}\,\eta$\cite{fischer2021}, where, $d_{10}=\braket{\psi_0\vert d(q)\vert\psi_1}=0.042\,ea_0$, is the fundamental transition dipole moment of the H-transfer system. Further, $\eta$ is a dimensionless function of the effective cavity volume and the dielectric constant within the cavity \cite{flick2017a,schaefer2018,fischer2021}, but treated here as a variable parameter chosen between $\eta=0.0$ (no molecule-cavity coupling) and $\eta=0.09$. The vibrational strong coupling (VSC) regime is determined by $0 < \eta < 0.1$, a definition in agreement with Ref.\cite{kockum2019}. Note, a proper lower bound on $\eta$ requires an energy relaxation channel, such that strong coupling is reached, when energy exchange between light-matter system constituents overcompensates energy relaxation to an external bath.\cite{mandal2022b}

Eventually, the dipole self-energy (DSE) reads 
\begin{equation}
\hat{H}_{DSE}
=
\dfrac{g^2_N}{\hbar\omega_c}
\sum^N_{i=1}
d^2(q_i)
+
\dfrac{g^2_N}{\hbar\omega_c}
\sum^N_{i\neq j}
d(q_i)\,
d(q_j)
\quad,
\label{eq.dipole_self_energy}
\end{equation}
containing both diagonal \textit{and} off-diagonal contributions, where the latter couples all $N$ H-transfer systems. For the $N$-molecule plus one-cavity mode system, an $N+1$-dimensional cavity potential energy surface (cPES) can be defined as
\begin{equation}
V_\eta(\underline{q},x_c)
=
V(\underline{q})
+
\dfrac{\omega^2_c}{2}
\biggl(
x_c
+
\sqrt{\dfrac{\hbar}{2\omega^3_c}}\,
g_N\,
d(\underline{q})
\biggr)^2
\quad.
\label{eq.cpes}
\end{equation}
For the uncoupled, single-molecule case ($\eta=0.0$, $N=1$), the cPES, $V(q,x_c)$, is shown in Figs.\ref{fig.htransfer_densities}b and \ref{fig.htransfer_densities}c superimposed with probability densities of the two lowest eigenfunctions, $\vert\Psi_0(q,x_c)\vert^2$ and $\vert\Psi_1(q,x_c)\vert^2$ obtained from diagonalizing the corresponding total Hamiltonian $\hat{H}$. These states are simple product states for $\eta=0.0$ and therefore also correspond to the enol and enethiol forms, weakly delocalized with small contributions at the other minimum as indicated. 

For $N$ molecules, in our fully quantum mechanical approach, the time-evolution of the light-matter hybrid system is governed by an $N+1$-dimensional time-dependent Schr\"odinger equation (TDSE)
\begin{equation}
\mathrm{i}\hbar
\dfrac{\partial}{\partial t}
\Psi(\underline{q},x_c,t)
=
\hat{H}\,
\Psi(\underline{q},x_c,t)
\quad,
\label{eq.tdse}
\end{equation}
which we solve numerically by means of the multiconfigurational time-dependent Hartree (MCTDH) approach\cite{beck2000,meyer2012} and its multilayer (ML-MCTDH) extension\cite{wang2003,manthe2008,vendrell2011} (\textit{cf.} SI, Sec. II for details) as implemented in the Heidelberg MCTDH package\cite{mctdh2019}. Moreover, we consider as initial state
\begin{equation}
\Psi_0(\underline{q},x_c)
=
\underbrace{\left(
\prod^N_{i=1}
\psi_\mathrm{OH}(q_i)
\right)}_{=\psi_N(\underline{q})}
\varphi_0(x_c)
\quad,
\label{eq.init_state}
\end{equation}
with ground state enol configuration, $\psi_\mathrm{OH}(q_i)$, for the $i$-th H-transfer system and cavity vacuum state, $\varphi_0(x_c)$. The latter corresponds to the ground state of the bare cavity with zero \textit{physical} photons. We will discuss differences to photon number expectation values for a molecule-cavity system at finite light-matter interaction strength ($\eta>0$) below. Finally, we note that $\Psi_0(\underline{q},x_c)$ turns out to be \textit{not} a good approximation to the vibrational polariton ground state under VSC in the herein discussed asymmetric transfer model, but leads to rich dynamics from where also H-transfer rates can be determined. A further analysis of the initial state is given in the SI, Sec. III.

\subsection{Observables}
We describe the time-evolution of the light-matter hybrid system by a time-dependent ensemble transfer probability from the enol form (the more stable configuration of the free molecule or the molecule in the cavity at $\eta=0.0$) to the enethiol form, which we define as
\begin{align}
P^\mathrm{ens}_\mathrm{SH}(t)
&=
\left\langle
\Psi(t)
\bigg\vert
\dfrac{1}{N}
\sum^N_{i=1}
\theta(q_i)
\bigg\vert
\Psi(t)
\right
\rangle
\quad.
\end{align}
Here, $\theta(q_i)$, is a Heaviside step function indicating a dividing surface located at the transition state of the individual transfer potentials. Due to the bound nature of the cPES, the transfer dynamics is subject to recrossing events at the dividing-surface, where we characterize the first recurrence by a \textit{recurrence time}, $\tau_r$. The latter allows us to introduce the notion of short-time dynamics for times, $t\leq \tau_r$, and subsequently the extraction of approximate short-time transfer rates from enol- to enethiol-configurations as
\begin{equation}
k^\mathrm{ens}_\mathrm{SH}
=
\frac{d}
{d t} 
P^\mathrm{ens}_\mathrm{SH}(t)
\bigg\vert_{t=t_\mathrm{max}}
\quad,
\end{equation} 
where $t_\mathrm{max}$ maximizes $k^\mathrm{ens}_\mathrm{SH}$ for $t_0<t_\mathrm{max}<\tau_r$. Further, time-dependent coordinate expectation values
\begin{equation}
\braket{R}(t)
=
\braket{
\Psi(t)
\vert
R
\vert
\Psi(t)
}
\quad,
\quad
R
=
q,x_c
\quad,
\end{equation}
provide a complementary perspective on the dynamics. In order to address cavity-induced collective quantum effects, we additionally study entanglement in the strongly coupled light-matter hybrid system \textit{via} von Neumann-entropies
\begin{equation}
S_i(t)
=
-
k_B\,
\mathrm{tr}
\{
\hat{\rho}_i(t)
\ln\hat{\rho}_i(t)
\}
\geq
0
\quad,
\label{eq.von_neumann_entropy}
\end{equation}
with Boltzmann constant, $k_B$, and reduced density operators, $\hat{\rho}_i(t)$, for an individual transfer mode, $i=q$, or the cavity mode, $i=C$. The equality in Eq.\eqref{eq.von_neumann_entropy} holds only if the reduced system is in a pure state, \textit{i.e.}, when the reduced subsystem is \textit{disentangled} from the remaining degrees of freedom. 

Finally, we consider the photon number expectation value, $\braket{\hat{n}_c}$, and its time evolution, which reads in length-gauge representation (\textit{cf.} {SI}, Sec. IV) 
\begin{equation}
\braket{\hat{n}_c}
=
\frac{1}{\hbar\omega_c}
\left(
\braket{\hat{H}_C}
+
\braket{\hat{H}_{SC}}
+
\braket{\hat{H}_{DSE}}
\right)
-
\dfrac{1}{2}
\quad.
\label{eq.length_gauge_number_operator}
\end{equation}
In the non-interacting limit, Eq.\eqref{eq.length_gauge_number_operator} reduces to, $\braket{\hat{n}_c}=\frac{1}{\hbar\omega_c}\braket{\hat{H}_C}-\frac{1}{2}=n_c$, with $n_c$ physical photons, whereas $n_c=0$ for the herein studied cavity vacuum state. For non-zero light-matter interaction, the photon expectation value initially reads, $\braket{\hat{n}_c}_0=\frac{1}{\hbar\omega_c}\left(\braket{\hat{H}_C}_0+\braket{\hat{H}_{DSE}}_0\right)-\frac{1}{2}>n_c$, due to a non-zero number of \textit{virtual} photons generated by the strong interaction of light and matter.\cite{kockum2019} In particular, the number of virtual photons at $t_0$ is directly determined by the DSE contribution and therefore ensemble size dependent (\textit{cf.} Eq.\eqref{eq.dipole_self_energy}). Note, the interaction term, $\braket{\hat{H}_{SC}}_0$, does initially not contribute to $\braket{\hat{n}_c}_0$ but will become relevant throughout the time-evolution of the hybrid system. 
\section{Results and discussion}
\label{sec3}
\subsection{Cavity-induced isomerization: Single molecule}
We start our discussion of cavity-induced isomerization for the asymmetric hydrogen transfer model in the single-molecule limit with $N=1$ by solving the TDSE \eqref{eq.tdse} for various coupling strengths, $\eta$, always using the same initial state \eqref{eq.init_state}. Tab.\ref{tab.OH_init_energies}, upper two lines, lists the corresponding initial state energies, $\braket{\hat{H}}_0$, and corresponding photon number expectation values, $\braket{\hat{n}_c}_0$, for selected values of $\eta$. We observe an increase for both expectation values, where $\braket{\hat{H}}_0>\Delta E^{cl}_\mathrm{OH}=1598\,\mathrm{cm}^{-1}$ for relatively strong couplings of $\eta>0.05$ and $\braket{\hat{n}_c}_0>0$ correspond to virtual photons generated by the DSE term. 
\begin{table}[hbt]
    \centering
    \begin{tabular}{l | r r r r r r }
       \hline\hline
         $\eta$ & \quad $0.00$ & \quad $0.01$ & \quad   $0.03$ & \quad   $0.05$  & \quad  $0.07$ & \quad $0.09$\\
       \hline
       $\braket{\hat{H}}_0/\,\mathrm{cm}^{-1}$ & $1098$ &  $1111$ & $1218$ & $1433$ & $1755$ & $2184$ \\
       $\braket{\hat{n}_c}_0$ & $0.0$ &  $0.16$ & $1.42$ & $3.95$ & $7.74$ & $12.80$ \\
       $\tau_r/\mathrm{fs}$  & -- & $87$ &  $88$ & $92$ & $95$ & $100$\\
       $k_\mathrm{SH}/10^{11}\,\mathrm{s}^{-1}$ & 0.00 & $0.02$ &  $0.15$ & $0.46$ & $0.95$ & $1.39$ \\
       \hline\hline
    \end{tabular}
\renewcommand{\baselinestretch}{1.}
\caption{Initial state energies, $\braket{\hat{H}}_0=\braket{\hat{H}_S}_0+\braket{\hat{H}_C}_0+\braket{\hat{H}_{DSE}}_0$, photon number expectation values, $\braket{\hat{n}_c}_0$, first-recurrence times, $\tau_r$, and short-time transfer rates, $k_\mathrm{SH}$, in the single-molecule limit for different light-matter interactions strengths, $\eta$. In all cases, the same initial state, $\Psi_0(q,x_c)=\psi_\mathrm{OH}(q)\,\varphi_0(x_c)$, was employed.
}
\label{tab.OH_init_energies} 
\end{table}

\begin{figure}[hbt]
\includegraphics[scale=1.0]{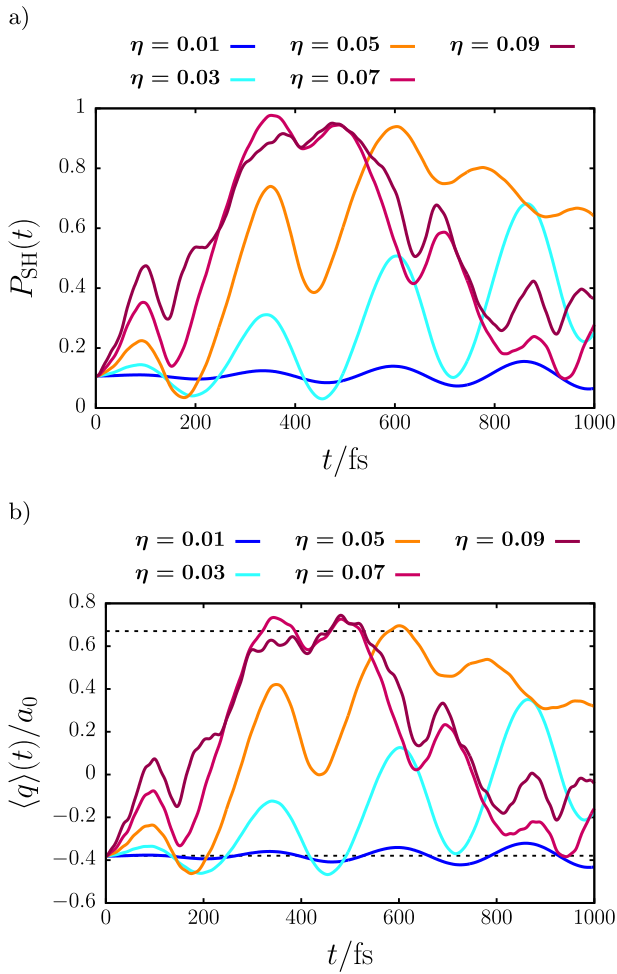}
\renewcommand{\baselinestretch}{1.}
\caption{Time-evolution of (a) single-molecule ($N=1$) transfer probability, $P_\mathrm{SH}(t)$, and (b) transfer coordinate expectation value, $\braket{q}(t)$, with black dashed lines indicating the quantum mechanical expectation values, $\braket{q}_\mathrm{OH}=\braket{\Psi_0\vert q\vert\Psi_0}$ and $\braket{q}_\mathrm{SH}=\braket{\Psi_1\vert q\vert\Psi_1}$, respectively, for different light-matter interaction strengths, $\eta$. }
\label{fig.SH_transfer}
\end{figure}

For $\eta>0$, H-transfer converting the enol (OH) to the enethiol (SH) form takes place. This can be seen from Fig.\ref{fig.SH_transfer}, where the transfer probability, $P_\mathrm{SH}(t)$ is shown (\ref{fig.SH_transfer}a), as well as the expectation value of the H-transfer coordinate, $\braket{q}(t)$ (\ref{fig.SH_transfer}b), both as a function of time and for different values of $\eta$. Note, for the transfer probability, $P_\mathrm{SH}(t)$, one initially finds $P_\mathrm{SH}(t_0)=0.11$ due to the weakly delocalized nature of $\psi_\mathrm{OH}(q)$. As time evolves, $P_\mathrm{SH}(t)$ increases for $\eta>0$ in an oscillatory fashion, which indicates formation of the enethiol isomer (Fig.\ref{fig.SH_transfer}a). Oscillatory signatures in $P_\mathrm{SH}(t)$ represent recurrences with a period of $264\,\mathrm{fs}$ for $\eta<0.07$, which resembles the cavity-mode energy, $\hbar\omega_c$. For stronger coupling, the dynamics turns out to be less regular. The transfer coordinate expectation value, $\braket{q}(t)$, closely resembles the transfer dynamics, with $\braket{q}<0$ indicating the enol and $\braket{q}>0$ the enethiol isomer (\textit{cf.} Fig.\ref{fig.SH_transfer}b).

From closer inspection of Fig.\ref{fig.SH_transfer}a, we can extract first-recurrence times, $\tau_r$, and corresponding short-time transfer rates, $k_\mathrm{SH}$, for different values of $\eta$. These are given in Tab.\ref{tab.OH_init_energies}, lower two rows. In the non-interacting limit $(\eta=0.0)$, we have $k_\mathrm{SH}=0.0$, \textit{i.e.}, there is \textit{no} population transfer to the local enethiol minimum without coupling to the cavity mode. In contrast, for $\eta>0$ we find transfer rates, $k_\mathrm{SH}\approx10^{9}\,\mathrm{s}^{-1}$ to $10^{11}\,\mathrm{s}^{-1}$, which increase with $\eta$ by nearly two orders of magnitude over the whole VSC regime between $\eta=0.01$ to $\eta=0.09$. The increase of reaction probability/\,transfer rate in a cavity for this particular system is in contrast to other systems, where a rate retardation has been found either experimentally\cite{ebbesen2016} or theoretically\cite{fischer2022}. That cavities can also enhance reactivity is a probably less widespread phenomenon, however, this possibility has been discussed in recent experimental\cite{ebbesen2021} and theoretical\cite{sun2022} work.

In order to interpret the positive effect of the cavity on the early-time, single-molecule transfer probability, $P_\mathrm{SH}(t)$, for TAA, we analyze the properties of the underlying single-molecule cPES, $V_\eta(q,x_c)$, which guides the dynamics of the vibro-polaritonic wave packet, $\Psi(q,x_c,t)$. 
\begin{figure*}[hbt!]
\includegraphics[scale=1.0]{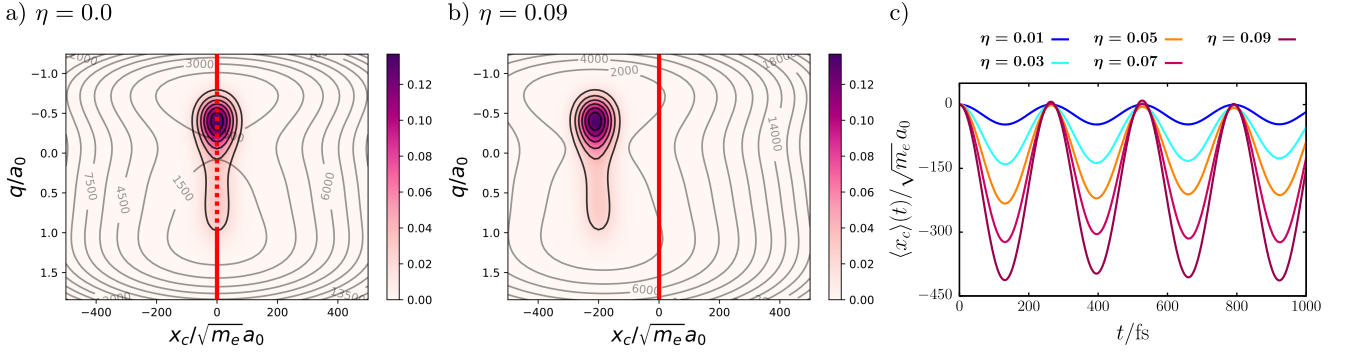}
\renewcommand{\baselinestretch}{1.}
\caption{Single-molecule cavity potential energy surface (cPES), $V_\eta(q,x_c)$, and vibro-polaritonic ground state densities, $\vert\Psi_0(q,x_c)\vert^2$, for different light-matter interaction strengths, $\eta=0.0$ (a), and $\eta=0.09$ (b), with initial cavity displacement coordinate expectation value, $\braket{x_c}_0=0$ indicated by red vertical line. (c) Time-evolution of cavity displacement coordinate expectation value, $\braket{x_c}(t)$, for different light-matter interaction strengths, $\eta$.}
\label{fig.cpes_coordexp_interpret}
\end{figure*}

In Figs.\ref{fig.cpes_coordexp_interpret}a and b, we show single-molecule cPES, $V_\eta(q,x_c)$, besides corresponding vibro-polaritonic ground state densities for different light-matter interaction strengths, with initial cavity displacement coordinate expectation value, $\braket{x_c}_0=0$, indicated by a red vertical line. For $\eta>0$, the cPES's minima are \textit{symmetrically} shifted to negative values of the cavity displacement coordinate, such that the cavity contribution of the initial wave packet naturally experiences an excitation. This is in contrast to a recently studied class of symmetric double well potentials, \textit{e.g.}, for the inversion of an NH$_3$ molecule\cite{fischer2022} or the ground state cPES of a cavity Shin-Metiu model\cite{li2021,sun2022}, which are asymmetrically distorted at finite light-matter interaction due to an antisymmetric, sign-changing dipole moment. The latter leads to barrier broadening, valley narrowing and (classical) dynamical caging effects, which in consequence reduce isomerization probabilities\cite{fischer2022,li2021}.

The static cPES perspective for TAA translates into a time-evolution of $\braket{x_c}(t)$ as shown in Fig.\ref{fig.cpes_coordexp_interpret}c. We find the vibro-polaritonic wave packet to acquire a significant dynamical component along the cavity displacement coordinate as time evolves due to the respective gradient on the cPES. $\braket{x_c}(t)$ reveals coherent oscillations with period $264\,\mathrm{fs}$ reflecting $\hbar\omega_c=\Delta \varepsilon_{10}$ and amplitude increasing with $\eta$, which resembles the enhanced cPES distortion in terms of altered turning points. Since the dynamics along cavity displacement and molecular transfer coordinates is naturally coupled \textit{via} the interaction term, $\hat{H}_{SC}$, we can interpret the isomerization as cavity-induced excitation along the transfer coordinate. The corresponding energy transfer can be related to a \textit{virtual} photon exchange between cavity and transfer modes, as will be discussed in detail below. Since the dynamics is strictly restricted to non-zero coupling strengths with $\eta>0.0$, the cavity can be interpreted as a ``catalyst'' in this model scenario -- despite the classical barrier height is not affected\cite{fischer2021}. We also note, the studied model system does not exhibit a ``reactant resonance effect'' as the local OH-/SH-stretching modes have frequencies, $\omega_\mathrm{OH}=3264\,\mathrm{cm}^{-1}$ and $\omega_\mathrm{SH}=2737\,\mathrm{cm}^{-1}$, which do not support localized bound states below the classical activation barriers.  

\subsection{Cavity-induced isomerization: Molecular ensembles}
We now extend our study to an ensemble of $N$ transfer systems coupled to a single cavity mode with initial state, $\Psi_0(\underline{q},x_c)$, given by Eq.\eqref{eq.init_state}. In what follows, we set $\eta=0.05$ and concentrate on the influence of varying ensemble sizes $N$ on the transfer process up to $N=20$. 
\begin{figure}[hbt!]
\includegraphics[scale=1.0]{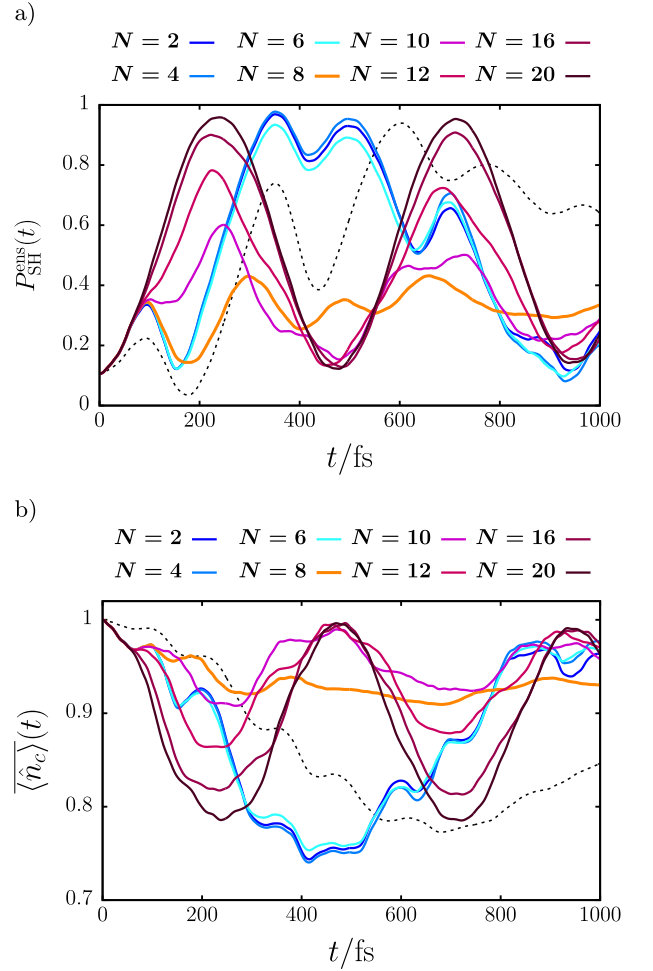}
\renewcommand{\baselinestretch}{1.}
\caption{Time-evolution of (a) ensemble transfer probability, $P^\mathrm{ens}_\mathrm{SH}(t)$, and (b) normalized photon number expectation value, $\overline{\braket{\hat{n}_c}}(t)$, as function of ensemble size $N$ for light-matter interaction strength, $\eta=0.05$. Single molecule properties ($N=1$) as reference indicated by black-dashed graphs.} 
\label{fig.SH_transfer_ensemble}
\end{figure}
At first, we discuss the time-evolution of ensemble transfer probabilities, $P^\mathrm{ens}_\mathrm{SH}(t)$, for different ensemble sizes $N$ as shown in Fig.\ref{fig.SH_transfer_ensemble}a. From the short-time dynamics, we extract an ensemble transfer rate, $k^\mathrm{ens}_\mathrm{SH}=3\times10^{12}\,\mathrm{s}^{-1}$, which is found to be two orders of magnitude larger than the single molecule rate ($0.46 \times10^{11}\,\mathrm{s}^{-1}$), and nearly independent of $N$ for ensemble sizes studied here. As time evolves, we observe an oscillatory evolution of $P^\mathrm{ens}_\mathrm{SH}(t)$, which can be classified by three different ``regimes'': 

(i) For $N\leq6$, the dynamics is dominated by a maximal probability density transfer at around $500\,\mathrm{fs}$ and $P^\mathrm{ens}_\mathrm{SH}(t)$ is modulated by a series of beats with varying amplitude and period of $264\,\mathrm{fs}$ corresponding to the cavity mode excitation energy of $\hbar\omega_c=126.5\,\mathrm{cm}^{-1}$. 

(ii) For $10\leq N\leq20$, two prominent maxima occur in $P^\mathrm{ens}_\mathrm{SH}(t)$ at around $200\,\mathrm{fs}$ and $700\,\mathrm{fs}$, with a significantly increased recurrence time of approximately $472\,\mathrm{fs}$, which we will again address below in context to entanglement of the cavity mode. 

(iii) Eventually, for an intermediate ensemble size with $N=8$, the probability transfer is significantly reduced (\textit{cf}. orange graph in Fig.\ref{fig.SH_transfer_ensemble}a) and no specific recurrence structure is observed.  

In order to provide an explanation for the ensemble transfer dynamics, we discuss a normalized photon number expectation value
\begin{equation}
\overline{\braket{\hat{n}_c}}(t)
=
\dfrac{\braket{\hat{n}_c}(t)}
{\braket{\hat{n}_c}(t_0)}
\quad,
\label{eq.relative_photon_nr}
\end{equation}
with, $\overline{\braket{\hat{n}_c}}(t_0)=1$, which allows us to address ensemble effects on the virtual photon transfer between cavity and molecules. We note, due to the different contributions to $\braket{\hat{n}_c}$ in Eq.\eqref{eq.length_gauge_number_operator}, a strict assignment of virtual photons only to the cavity mode is in principle not possible as both interaction and DSE term also contribute significantly to $\braket{\hat{n}_c}(t)$. The time-evolution of $\overline{\braket{\hat{n}_c}}(t)$ for different $N$ is shown in Fig.\ref{fig.SH_transfer_ensemble}b and we find $\overline{\braket{\hat{n}_c}}(t)<1$ for all $N$ (including $N=1$) over the studied time-interval, \textit{i.e.}, virtual photons are transferred to the molecular ensemble. In particular, we observe $\overline{\braket{\hat{n}_c}}(t)$ to qualitatively resemble the \textit{inverse} dynamical trend in $P^\mathrm{ens}_\mathrm{SH}(t)$, \textit{i.e.}, virtual photon transfer to the molecular ensemble coincides with enhanced population transfer to the enethiol region (\textit{cf.} Fig.\ref{fig.SH_transfer_ensemble}a). Hence, virtual photons are not only exchanged with the transfer ensemble but virtually drive the cavity-induced isomerization. We note, \textit{non-normalized} expectation values, $\braket{\hat{n}_c}$, significantly depend on the interaction regime,\textit{i.e.}, $\eta$ (\textit{cf.} {SI}). 
\subsection{Cavity-induced entanglement in ensembles}
In order to gain further insight, we address differences in entanglement between the single-molecule limit and the ensemble scenario, which allows us to formulate an interpretation for ensemble-enhanced isomerization. We first realize, that in the ensemble scenario, an enhanced \textit{ensemble} transfer rate cannot straightforwardly be explained by the distortion of the ($N+1$-dimensional) cPES, since the distortion in a single molecule-cavity subspace decreases due to the Dicke-type light-matter interaction used here, $g_N$, as, $g_N\sim \frac{1}{\sqrt{N}}$, for increasing $N$ (\textit{cf.} Eq.\eqref{eq.cpes}).\cite{sun2022} However, due to strong coupling between light and matter constituents, entanglement effects are in contrast expected to shape the \textit{ensemble} transfer dynamics. 
\begin{figure}[hbt!]
\includegraphics[scale=1.0]{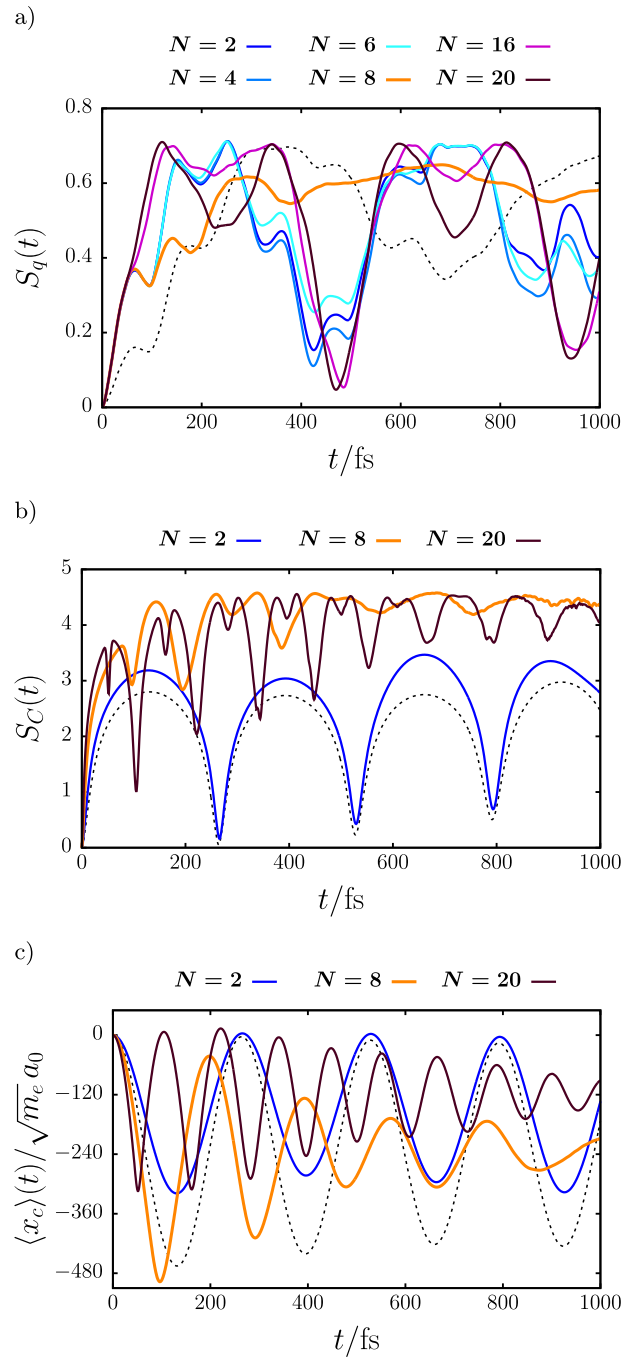}
\renewcommand{\baselinestretch}{1.}
\caption{Time-evolution of (a) transfer von Neumann-entropy, $S_q(t)$, (b) cavity von Neumann-entropy, $S_C(t)$, and (c) cavity displacement coordinate expectation value, $\braket{x_c}(t)$, as function of ensemble size $N$ for light-matter interaction strength, $\eta=0.05$. Single molecule properties ($N=1$) as reference indicated by black-dashed graphs.} 
\label{fig.entropies_transfer_ensemble}
\end{figure}

We quantify entanglement by transfer and cavity von Neumann-entropies, $S_q(t)$, and $S_C(t)$, as defined in Eq.\eqref{eq.von_neumann_entropy} with time-evolution depicted in Figs.\ref{fig.entropies_transfer_ensemble}a and b for different ensemble sizes, $N$. Initially, we have, $S_q(t_0)=S_C(t_0)=0$, independent of ensemble size due to the product form of the initial state in Eq.\eqref{eq.init_state}, which is by definition disentangled. From a wave function perspective, build up of entanglement can therefore be interpreted in terms of an increase of the multiconfigurational character of the full vibro-polaritonic wave function, \textit{i.e.}, as deviation from a \textit{disentangled} product state. In general, we find the time-evolution of both entropies to differ significantly due to the different nature of both subsystems and, $S_C(t)>S_q(t)$. 

Starting with the reduced transfer system perspective, we observe $S_q(t)$ to increase faster with time for increasing $N$. This indicates a faster build-up of subsystem entanglement or equivalently multiconfigurational character in the vibro-polaritonic wave function with respect to H-transfer systems. Further, for $N>1$, we observe an increase in $S_q(t)$ to accompany the transfer process to the enethiol configuration as characterized by $P^\mathrm{ens}_\mathrm{SH}(t)$ (\textit{cf.} Fig.\ref{fig.SH_transfer_ensemble}a), \textit{i.e.}, H-transfer relates to enhanced entanglement. Further, the transfer von Neumann-entropy reaches (local) minima, \textit{i.e.}, reduced transfer system entanglement, where extrema are observed in the ensemble transfer probability. This observation is in line with above discussed dynamics of $\overline{\braket{\hat{n}_c}}(t)$ (\textit{cf.} Fig.\ref{fig.SH_transfer_ensemble}b), \textit{i.e.}, virtual photon transfer drives isomerization, which naturally leads to a stronger entanglement (larger $S_q(t)$) between cavity and transfer systems. In contrast, suppressed transfer relates to small values of $S_q(t)$. Moreover, for intermediate ensemble sizes of $N=8$, $S_q(t)$ differs significantly by exhibiting a nearly monotonic but slow increase with time, which is accompanied by suppressed virtual photon transfer and isomerization probability (\textit{cf.} Fig.\ref{fig.SH_transfer_ensemble}), in line with the previous argument. 

Turning to the cavity mode, we find $S_C(t)$ to be characterized by an overall increase with time, which becomes strongly pronounced for large $N$, \textit{i.e.}, the cavity mode becomes quickly strongly entangled with the transfer ensemble. Further, the cavity von Neumann-entropy mimics the vibro-polaritonic wave packet's dynamics along the cavity displacement coordinate as captured by $\braket{x_c}(t)$ (\textit{cf.} Fig.\ref{fig.entropies_transfer_ensemble}c). For small ensembles ($N=2$), $S_C(t)$ oscillates with a period of $264\,\mathrm{fs}$, recovering the  cavity mode frequency and is minimized for $\braket{x_c}(t)$ reaching the initial position at $x_c=0$. In contrast, for large ensembles with $N=20$, the cavity entropy exhibits minima characterized by two different time scales, which can be related to both maximal and minimal displacements of the vibro-polaritonic wave packet along the displacement coordinate $x_c$. Eventually, we note two characteristics, which additionally indicate the complex nature of the ensemble dynamics: First, for increasing $N$ the amplitude of $\braket{x_c}(t)$ is damped as time evolves since the cavity mode is increasingly immersed in a \textit{bath} of strongly coupled and highly anharmonic transfer systems. Second, as ensemble size $N$ increases, the initial amplitude of $\braket{x_c}(t)$ increases too with a maximum around $N=8$. We attribute this transition to a change from a small transfer ensemble of multi-mode Rabi type to a ``large'' ensemble resembling a system-bath-type regime with a cavity mode subject to dissipation on the time scale shown. Naturally, this transition is equivalently observed in the transfer dynamics.

In summary, we attribute isomerization in the model studied here to be induced by virtual photon transfer, which maximizes time-dependent \textit{changes} in transfer system entanglement quantified by $S_q(t)$. In particular, ensemble-enhanced isomerization rates relate to a cavity-induced entanglement effect between transfer systems, which is not explainable by classical cPES distortion arguments valid in the single-molecule limit. Accordingly, the herein discussed cavity-induced ensemble transfer process can be interpreted as an inherently collective quantum mechanical effect, which in particular cannot be captured by scaled single-molecule models.

\section{Summary and conclusions}
\label{sec4}
In this work, we studied the quantum dynamics of an entangled molecular ensemble for an asymmetric hydrogen transfer model of thioacetylacetone (TAA) interacting with a single cavity mode under vibrational strong coupling. An $N$-molecule form of the Pauli-Fierz Hamiltonian was used beyond frequently adopted model Hamiltonians such as the Tavis-Cummings or Dicke-models. A $N+1$-dimensional time-dependent Schr\"odinger equation was solved numerically by means of the MCTDH ansatz to follow the cavity-induced isomerization dynamics from enol to enethiol isomers of TAA.
  
At finite light-matter interaction, the cavity acts as a ``catalyst'' by inducing population transfer to the enethiol isomer, which is energetically less favorable than the enol form of TAA outside the cavity or at vanishing cavity-molecule coupling. This process is identified to be driven by virtual photon transfer to the $N$-molecule subsystem. We extract approximate short-time transfer rates, which span in the single-molecule limit two orders of magnitude for increasing light-matter interaction. In an entangled ensemble of transfer systems, with a collective Dicke-type light-matter coupling, a collectively enhanced transfer rate is observed following from an interplay of virtual photon-transfer and non-trivial entanglement dynamics between light and matter components of the hybrid system. We furthermore find non-trivial ensemble size dependence of the dynamics as $N$ grows, which we attribute to a transition from a multi-mode Rabi type scenario to a system-bath-type regime, where the \textit{cavity mode} is effectively immersed in a bath of strongly coupled, anharmonic transfer systems. Our study points at the highly non-trivial role of quantum effects in molecular ensemble models strongly interacting with a quantized cavity mode beyond scaled single-molecule dynamics.

We close by pointing out several possible extensions of our model. First, we do not take into account dissipative effects and decoherence due to leaky cavity modes or the presence of other molecular degrees of freedom, which will naturally influence the transfer probability and allow for a more rigorous definition of transfer rates. However, if we assume additional degrees of freedom to be only weakly coupled, which is relevant to ensure the VSC regime, the main findings of our work should qualitatively remain the same for the herein studied time-interval. Further, we did not take into account finite temperature effects, which can be assumed relevant due to the relatively low energy scale in our model. Finally, while we went well beyond the Dicke-model by using an ensemble formulation based on the Pauli-Fierz Hamiltonian, we still assumed a Dicke-type coupling in our work. It might be instructive to also lift the Dicke-type perspective for the molecule-cavity coupling to carefully discuss deviations and potentially emerging properties in vibro-polaritonic chemistry for less restricted coupling models. Along similar lines, orientational effects (due to rotation of molecules) on the molecule-cavity coupling in fluctuating molecular ensembles, their influence on the related entanglement dynamics as well as the inclusion of direct intermolecular interactions could be interesting milestones towards a realistic description of molecular ensembles in cavities.
\section*{Acknowledgements}
We acknowledge fruitful discussions with Oliver K\"uhn (Rostock) and Foudhil Bouakline (Potsdam). This work was funded by the Deutsche Forschungsgemeinschaft (DFG, German Research Foundation) under Germany's Excellence Strategy -- EXC 2008/1-390540038. E.W. Fischer acknowledges support by the International Max Planck Research School for Elementary Processes in Physical Chemistry.

\section*{Data Availability Statement}
The data that support the findings of this study are available from the corresponding author upon reasonable request.

\section*{Conflict of Interest}
The authors have no conflicts to disclose.

\end{document}


\title{Supplementary Information: 
\\
 Cavity-Catalyzed Hydrogen Transfer Dynamics in an Entangled Molecular Ensemble under Vibrational Strong Coupling}
\author{Eric W. Fischer}
\email{ericwfischer.sci@posteo.de}
\affiliation{Theoretische Chemie, Institut f\"ur Chemie, Universit\"at Potsdam,
Karl-Liebknecht-Strasse 24-25, D-14476 Potsdam-Golm, Germany}

\author{Peter Saalfrank}
\affiliation{Theoretische Chemie, Institut f\"ur Chemie, Universit\"at Potsdam,
Karl-Liebknecht-Strasse 24-25, D-14476 Potsdam-Golm, Germany}
\affiliation{Institut f\"ur Physik und Astronomie, Universit\"at Potsdam, Karl-Liebknecht-Stra\ss e 24-25, D-14476 Potsdam-Golm, Germany}
\let\newpage\relax

\begin{abstract}
We provide supplementary information concerning (1) the derivation of the one-dimensional hydrogen transfer Hamiltonian and deviations from a reaction path Hamiltonian, (2) numerical details on quantum dynamics calculations via the (ML)MCTDH methods, (3) details on the initial state in terms of the vibro-polaritonic density of states and (4) a derivation of the photon number operator in length-gauge representation.
\end{abstract}

\let\newpage\relax
\maketitle
\newpage

\setcounter{equation}{0}
\renewcommand{\theequation}{\thesection\arabic{equation}}
\renewcommand{\thefigure}{S\arabic{figure}}
\section{One-dimensional Hydrogen Transfer Reaction Hamiltonian}
\subsection{Reaction Potential and Minimum Energy Path}
We derive the one-dimensional H-transfer Hamiltonian, $\hat{H}_S$ (Eq.(2) with $N=1$ in the main text ), from a two-dimensional asymmetric H-transfer reaction Hamiltonian for thioacetylacetone (TAA) developed by Dosli{\'c} \textit{et al.}\cite{doslic1999}. This Hamiltonian was constructed from {\em ab initio} electronic structure calculations and reads 
\begin{equation}
\hat{H}_R
=
-
\dfrac{\hbar^2}{2\mu_S}\dfrac{\partial^2}{\partial q^2}
-
\dfrac{\hbar^2}{2\mu_B}\dfrac{\partial^2}{\partial Q^2}
+
V(q,Q)
\quad,
\label{eq.2d_htransfer_hamiltonian}
\end{equation}
with a (H-transfer) reaction coordinate, $q$, a (collective) ``heavy'' mode coordinate, $Q$, and corresponding reduced masses, $\mu_S=1914.028\,m_e$ and $\mu_B=8622.241\,m_e$, respectively.\cite{doslic1999} The two-dimensional molecular potential energy surface (PES), $V(q,Q)$, is given by
\begin{equation}
V(q,Q)
=
V(q)
+
\dfrac{\mu_B\,\omega^2_B}{2}
\left(
Q
-
\lambda_S(q)
\right)^2
\quad,
\label{eq.2d_htransfer_pes}
\end{equation}
with ``heavy'' mode frequency, $\omega_B=0.0009728\,E_h$, and nonlinear coupling function, $\lambda_S(q)=a_S\,q^2+b_S\,q^3$, determined by parameters, $a_S=0.794\,a^{-1}_0$ and $b_S=-0.2688\,a^{-2}_0$. The reaction path potential is described in terms of an adiabatic potential
\begin{equation}
V(q)
=
\dfrac{1}{2}
\biggl(
V_+(q)
-
\sqrt{
V^2_-(q)
+
4\,K^2(q)}
\biggr)
\quad,
\end{equation}
where, $V_\pm(q)=V_1(q)\pm V_2(q)$, with diabatic harmonic PES, $V_i(q)$, and non-adiabatic coupling function, $K(q)$, defined as
\begin{align}
V_i(q)
&=
\dfrac{\mu_i\,\omega^2_i}{2}\,(q-q_{i,0})^2
+
\Delta_i
\quad,
\vspace{0.2cm}
\\
K(q)
&=
k_c\,\exp\biggl(-(q-q_c)^2\biggr)
\quad.
\end{align}
The harmonic potentials resemble the R--OH ($V_1(q)$) and R--SH ($V_2(q)$) configurations in TAA with corresponding harmonic frequencies, $\omega_\mathrm{OH}=0.01487\,E_h/\hbar$ and $\omega_\mathrm{SH}=0.01247\,E_h/\hbar$, reduced masses, $\mu_\mathrm{OH}=1728.46\,m_e$ and $\mu_\mathrm{SH}=1781.32\,m_e$, relative energy shifts, $\Delta_\mathrm{OH}=0.0\,E_h$ and $\Delta_\mathrm{SH}=0.003583\,E_h$, as well as displacements, $q_{\mathrm{OH},0}=-0.7181\,a_0$ and $q_{\mathrm{SH},0}=1.2094\,a_0$. The coupling function, $K(q)$, is determined by an amplitude, $k_c=0.15582\,E_h$, and a displacement, $q_c=0.2872\,a_0$.\cite{doslic1999} Further, a molecular dipole function (neglecting the vector character of the dipole moment) is given in Ref.\cite{doslic1999} as 
\begin{multline}
d(q,Q)
=
d_0
+
d_S(q-q_0)
+
d_B(Q-\lambda_S(q))
\\
+
d_{SB}(q-q_0)(Q-\lambda(q))
\quad,
\end{multline}
with parameters, $d_0=1.68\,ea_0$, $d_S=-0.129\,ea_0/a_0$, $d_B=0.023\,ea_0/a_0$, $d_{SB}=0.451\,ea_0/a^2_0$ and $q_0=-0.59\,a_0$. 
\begin{figure*}[hbt!]
\includegraphics[scale=1.0]{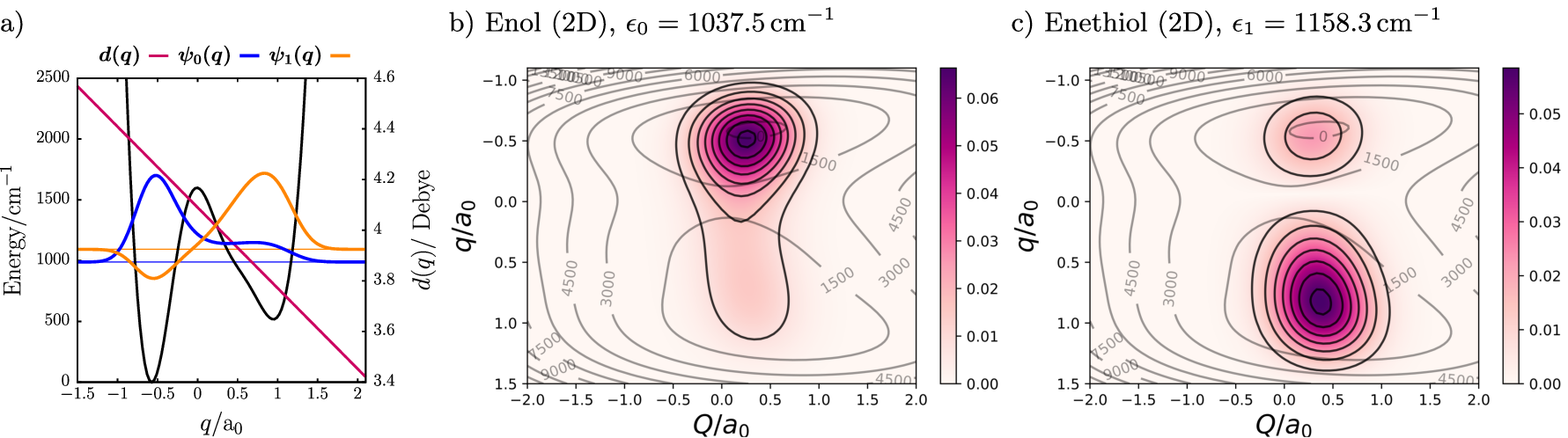}
\renewcommand{\baselinestretch}{1.}
\caption{(a) One-dimensional hydrogen-transfer reaction potential, $V(q)$ (in black), with dipole function, $d(q)$, and two lowest eigenstates, $\psi_0(q)=\psi_{\mathrm{OH}}(q)$ and $\psi_1(q)=\psi_{\mathrm{SH}}(q)$. (b) Ground state, $\vert\psi_0(q,Q)\vert^2=\vert\psi_{\mathrm{OH}}(q,Q)\vert^2$, and (c) first excited state densities, $\vert\psi_1(q,Q)\vert^2=\vert\psi_{\mathrm{SH}}(q,Q)\vert^2$, of two-dimensional reaction Hamiltonian, $\hat{H}_R$, in Eq.\eqref{eq.2d_htransfer_hamiltonian} embedded in two-dimensional molecular PES, $V(q,Q)$, given in Eq.\eqref{eq.2d_htransfer_pes} with contours in $\mathrm{cm}^{-1}$.} 
\label{fig.htransfer_1D}
\end{figure*}

In the present work, where we focus on the ensemble character of isomerizing molecules, an effective approximate one-dimensional Hamiltonian, $\Hat{H}_S$, and corresponding dipole function, $d(q)$, are constructed, which still resemble the main features of their two-dimensional counterparts. We derive the one-dimensional transfer Hamiltonian by minimizing, $V(q,Q)$, with respect to $Q$ as
\begin{equation}
\dfrac{\partial}{\partial Q}V(q,Q)
=
0
\quad
\Leftrightarrow
\quad
Q_\mathrm{0}
=
\lambda(q)
\quad,
\label{eq.min_bath_coord}
\end{equation}
such that the transfer potential and the dipole function subsequently simplify to one-dimensional functions
\begin{align}
V(q,Q_0)
&=
V(q)
\quad,
\vspace{0.2cm}
\\
d(q,Q_0)
&=
d(q)
= 
d_0
+
d_S
(q-q_0)
\quad.
\end{align}
The latter holds equivalently for an ensemble of $N$ transfer ensembles. In our study, we neglect the ``heavy mode'', $Q$, which does not couple \textit{via} a potential-like term to the transfer coordinate, $q$. In Fig.\ref{fig.htransfer_1D}, we show $V(q)$ and $d(q)$, with the lowest two eigenfunctions, $\psi_0(q)=\psi_{\mathrm{OH}}(q)$ (enol) and $\psi_1(q)=\psi_{\mathrm{SH}}(q)$ (enethiol), indicated. The latter were obtained by diagonalizing $\hat{H}_S$ in terms of a Colbert-Miller discrete variable representation (DVR)\cite{colbert1992} for the transfer coordinate with $N_q=121$ grid points and $q\in[-1.5,2.1]\,a_0$. The corresponding eigenenergies are $\varepsilon_0=988.3\,\mathrm{cm}^{-1}$ and $\varepsilon_1=1092.8\,\mathrm{cm}^{-1}$ as stated in the main text with an energy difference of $\Delta \varepsilon_{10}=\varepsilon_1-\varepsilon_0=126.5\mathrm{cm}^{-1}$. 

For the two-dimensional Hamiltonian, $\hat{H}_R$, in Eq.\eqref{eq.2d_htransfer_hamiltonian}, we numerically obtain energies, $\epsilon_0=1037.5\,\mathrm{cm}^{-1}$ and $\epsilon_1=1158.3\,\mathrm{cm}^{-1}$, for the ground and first excited states, respectively, with energy difference of $\Delta \epsilon_{10}=120.8\,\mathrm{cm}^{-1}$. Here, were we again employed a Colbert-Miller DVR with transfer grid parameters equivalent to the one-dimensional case discussed above and ``heavy'' mode coordinate $Q\in[-2.0,2.0]\,a_0$ with $N_Q=61$ grid points. Eventually, classical activation energies are by construction equivalent for the one- and two-dimensional PES with $\Delta E^{cl}_\mathrm{OH}=1598\,\mathrm{cm}^{-1}$ and $\Delta E^{cl}_\mathrm{SH}=1081\,\mathrm{cm}^{-1}$ as stated in the main text, since $V(q)$ is equivalent to the reaction potential along the minimum energy path on $V(q,Q)$. 

\subsection{Deviations from a Reaction Path Hamiltonian}
We discuss deviations of our approach from a reaction path Hamiltonian, which additionally involves kinetic energy couplings due to non-zero reaction path curvature. Miller, Handy and Adams\cite{miller1979} showed that a reaction path Hamiltonian of a two-dimensional system with mass-weighted, cartesian-like coordinates is given by
\begin{equation}
\hat{H}(\hat{p}_s,s,\hat{P}_s,Q_s)
=
\dfrac{\hat{p}^2_s}{2\left(1+Q_s\,\kappa(s)\right)^2}
+
V_0(s)
+
\hat{H}_\mathrm{valley}(s)
\quad,
\end{equation} 
where the first two terms correspond to kinetic and potential energy contributions along the reaction coordinate, $s$, with conjugate momentum, $\hat{p}_s$, whereas the third term provides the ``valley'' Hamiltonian
\begin{align}
\hat{H}_\mathrm{valley}(s)
&=
\frac{\hat{P}^2_s}{2}
+
\frac{\omega(s)^2}{2}\,Q^2_s
\quad.
\end{align}
The latter accounts for a $s$-dependent ``valley'' mode with frequency, $\omega(s)$, perpendicular to the reaction path that couples to the reaction coordinate {\em via} the reaction path curvature, $\kappa(s)$. For the hydrogen transfer system studied here, we have by construction, $V_0(s)=V(q,Q_0)=V(q)$. 
\begin{figure}[hbt!]
\includegraphics[scale=1.0]{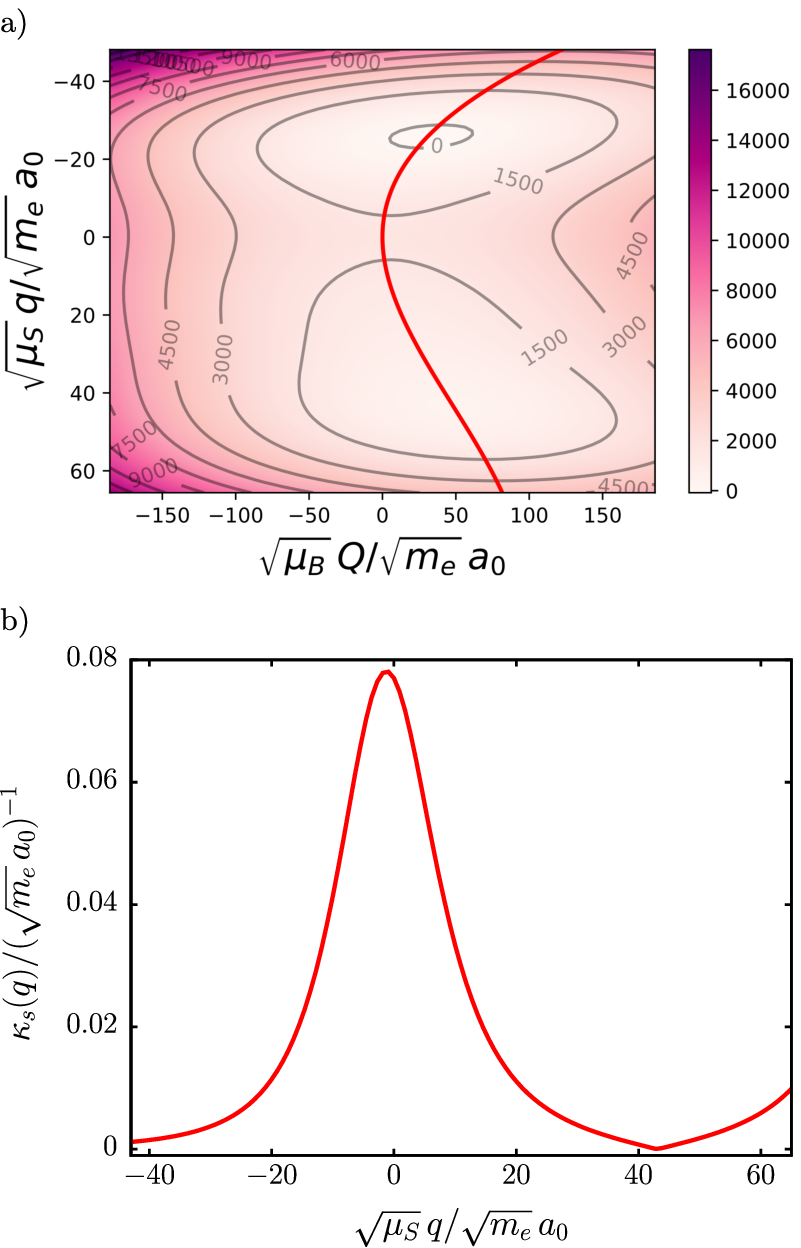}
\renewcommand{\baselinestretch}{1.}
\caption{a) Contour plot of molecular PES, $V(q,Q)$, in mass-weighted coordinates with reaction path, $\underline{s}(q)$, in red and colorbar in wave numbers $(\mathrm{cm}^{-1})$ and b) minimum energy path curvature, $\kappa_s(q)$, parameterized by mass-weighted transfer coordinate, $\sqrt{\mu_S}\,q$ with maximum at the transition state.} 
\label{fig.reactionpath_curvature}
\end{figure}

In the following, we discuss deviations from $\hat{H}(\hat{p}_s,s,\hat{P}_s,Q_s)$, which emerge when approximating the reaction path contribution by the bare transfer Hamiltonian 
\begin{equation}
\hat{H}_S
=
\dfrac{\hat{p}^2_q}{2}
+
V(q)
=
-
\dfrac{\hbar^2}{2\mu_S}
\dfrac{\partial^2}{\partial q^2}
+
V(q)
\quad.
\end{equation}
This assumption is equivalent to approximately decoupling the ``valley'' Hamiltonian, $\hat{H}_\mathrm{valley}(s)$, from the reaction path contribution by assuming the reaction path curvature, $\kappa(s)$, to be small. In order to access this condition, we discuss the curvature, $\kappa(s)$, of the minimum energy or reaction path, $\underline{s}(q)$, which we introduce as parametric curve in the mass-weighted $q$-$Q$-plane\cite{fischer2022} 
\begin{equation}
\underline{s}(q)
=
\left(s_1(q),s_2(q)\right)^T
=
\left(\sqrt{\mu_S}\,q,\sqrt{\mu_B}\,Q_0(q)\right)^T
\quad,
\end{equation} 
with components, $s_1(q)$ and $s_2(q)$. Here, $Q_0(q)=\lambda_S(q)$, as derived in Eq.\eqref{eq.min_bath_coord}, which minimizes, $V(q,Q)$, with respect to variations in the ``heavy'' mode coordinate. From $\underline{s}(q)$, which is parameterized in terms of the hydrogen-transfer coordinate, $q$, we obtain the corresponding curvature, $\kappa_s(q)$, as\cite{arens2018}
\begin{equation}
\kappa_s(q)
=
\dfrac{\mathrm{\det}\left(\underline{s}^\prime,\underline{s}^{\prime\prime}\right)}{\vert\vert \underline{s}^\prime\vert\vert^3}
=
\dfrac{\vert s^\prime_1 s^{\prime\prime}_2-s^{\prime\prime}_1 s^\prime_2\vert}{\left[(s^\prime_1)^2+(s^\prime_2)^2\right]^{\frac{3}{2}}}
\quad,
\end{equation}
with derivatives, $s^\prime_i=\frac{\partial}{\partial q}s_i(q)$ and $s^{\prime\prime}_i=\frac{\partial^2}{\partial q^2}s_i(q)$, respectively. We like to emphasize, that $\kappa_s(q)$ depends now on the hydrogen transfer coordinate, $q$, which parametrizes the reaction path. Further, for reaction path elements, $s_1(q)=\sqrt{\mu_S}\,q$ and $s_2(q)=\sqrt{\mu_B}\,\lambda_S(q)$, we find derivatives
\begin{align}
s^\prime_1
&=
\sqrt{\mu_S}
\quad,
\vspace{0.2cm}
\\
s^\prime_2(q)
&=
\sqrt{\mu_B}
\left(
2\,a_S\,q
+
3\,b_S\,q^2
\right)
\quad,
\end{align}
and
\begin{align}
s^{\prime\prime}_1
&=
0
\quad,
\vspace{0.2cm}
\\
s^{\prime\prime}_2(q)
&=
\sqrt{\mu_B}
\left(
2\,a_S
+
6\,b_S\,q
\right)
\quad,
\end{align}
which allow us to write the curvature explicitly as
\begin{equation}
\kappa_s(q)
=
\dfrac{2\sqrt{\mu_S\,\mu_B}\,\vert a_S+3\,b_S\,q\vert}
{\left[\mu_B\,q^2\left(2\,a_S+3\,b_S\,q\right)^2+\mu_S\right]^{\frac{3}{2}}}
\quad.
\end{equation}
In Fig.\ref{fig.reactionpath_curvature}a, we show the two-dimensional molecular PES, $V(q,Q)$, with reaction path, $\underline{s}(q)$, in red and the corresponding curvature, $\kappa_s(q)$, in Fig.\ref{fig.reactionpath_curvature}b. A kinetic separation of the reaction path from the ``valley'' coordinate is a good approximation, if
\begin{equation}
\hat{T}_s
=
\dfrac{\hat{p}^2_s}{2\left(1+Q_s\,\kappa_s(q)\right)^2}
\approx
\dfrac{\hat{p}^2_q}{2}
=
\hat{T}_q
\quad,
\end{equation}
which holds for $\vert Q_s\,\kappa_s(q)\vert \ll 1$. By taking into account the maximal curvature, $\kappa_s(q=0.0)\approx0.078\,(\sqrt{m_e}\,a_0)^{-1}$, at the transition state ($q=0.0$), the coupling is solely determined by the ``valley'' coordinate's magnitude, which can be traced back to the excitation of the two-dimensional transfer system along the ``heavy'' mode. We shall estimate the latter by means of of the harmonic ``valley'' potential's classical turning points
\begin{equation}
Q^\pm_s
=
\pm \sqrt{\frac{\hbar}{\omega_B}(2v+1)}
\quad,
\end{equation}
with vibrational quantum number, $v$, and, $\omega(q)=\omega_B$, at the transition state. For, $v=1$, we find, $\vert Q^\pm_s\,\kappa_s(q=0.0)\vert \approx 4.3$, \textit{i.e.}, in principal already for the ``heavy'' mode being excited to the first excited state, which has to be expected during the transfer process, we observe a coupling to the reaction coordinate that is assumed to alter the transfer dynamics of the molecular isomerization model system. However, as the role of the ``heavy'' mode is \textit{not} central for the \textit{cavity-induced} isomerization dynamics, we consider our approximation to be qualitatively valid and sufficient to discuss entanglement-induced collective effects in the herein studied reactive vibro-polaritonic model.

\section{Numerical details for Quantum Dynamics}
\setcounter{equation}{0}
\begin{figure*}[hbt!]
\includegraphics[scale=1.0]{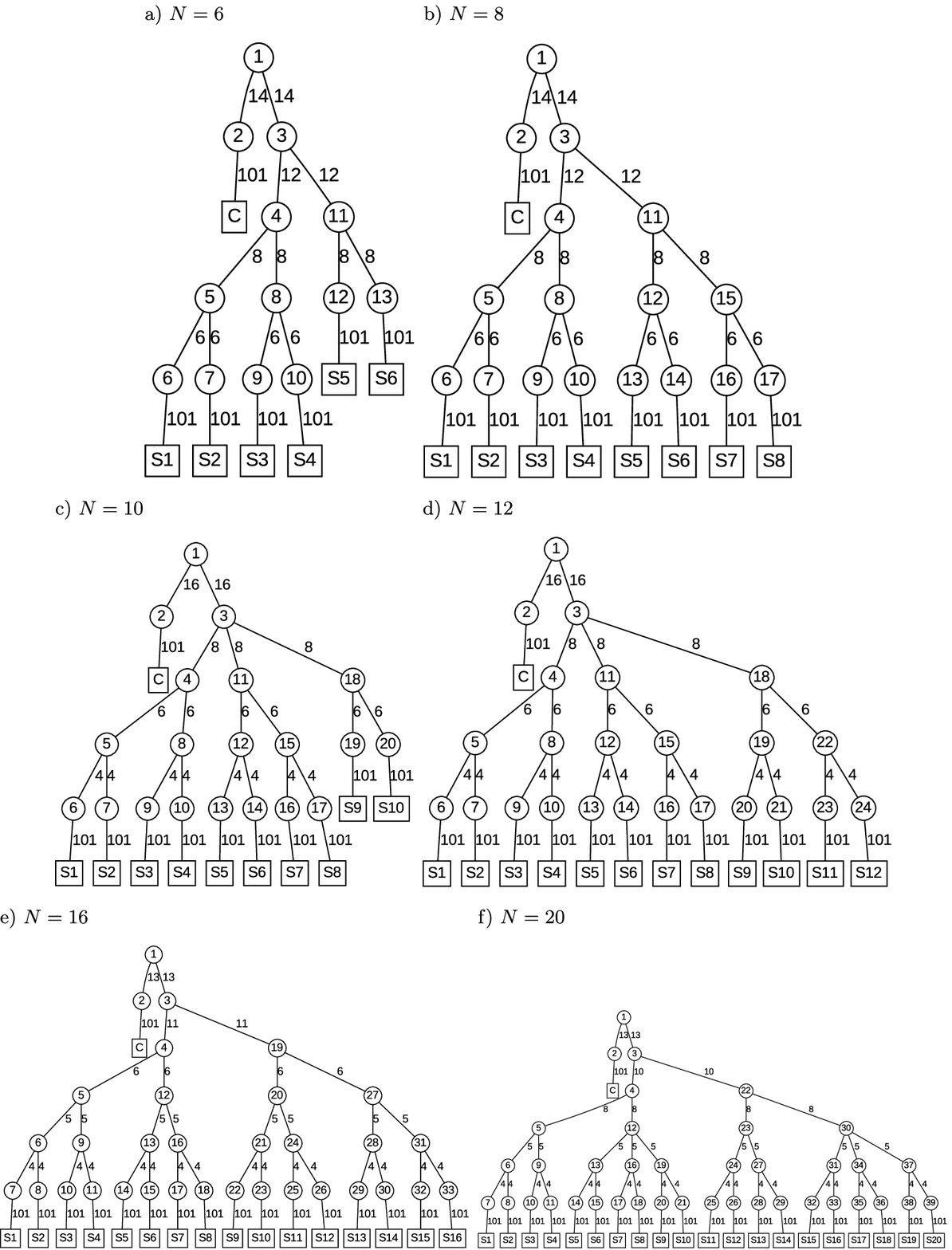}
\renewcommand{\baselinestretch}{1.}
\caption{Multilayer trees for different ensemble sizes $N$ with $S1$ to $SN$ transfer systems and cavity mode $C$. Number of SPFs are shown next to bonds connecting circular nodes and number of primitive basis functions are shown next to bonds connecting circular and square nodes. All trees are employed for light-matter interaction $\eta=0.05$.} 
\label{fig.ml_trees_ensemble}
\end{figure*}
We solve the TDSE (Eq.(8) in the main text) numerically by means of the multiconfigurational time-dependent Hartree (MCTDH) method and its multilayer extension (ML-MCTDH). We propagate up to final time, $t_f=1000\,\mathrm{fs}$, unless stated otherwise, and employ a Colbert-Miller DVR\cite{colbert1992} for transfer reaction coordinates, $q_i\in[-1.5,2.1]a_0$, with $N_q=101$ grid points as well as a harmonic oscillator (HO) DVR for the cavity mode with $N_c=101$ grid points and $x_c\in[-561.35,+561.35]\sqrt{m_e}\,a_0$. We treat ensembles up to $N=4$ via the MCTDH method with single particle functions (SPFs), $n_s=n_c=10$. For ensembles with $4<N\leq 20$, we employ the ML-MCTDH method with converged trees (max. natural population $\leq10^{-4}$) for all $N$ as displayed in Fig.\ref{fig.ml_trees_ensemble}. We employ the same DVR with identical number of primitive basis functions as above independent of ensemble size, $N$, but $N$-dependent numbers of SPFs, as shown next to bonds in ML trees. The latter is a result of different entanglement structure in the full vibro-polaritonic wave packet for different $N$.

\section{Initial State and Vibro-Polaritonic Density of States}
\setcounter{equation}{0}
We analyze the initial state (\textit{cf.} Eq.(9) in main text) in terms of its vibro-polaritonic contributions, which we access by means of the vibro-polaritonic density of states (DOS) 
\begin{align}
\sigma(\hbar\omega)
&=
\int^{2t_f}_0
C(t)\,
e^{\mathrm{i}E\,t/\hbar}
\mathrm{d}t
\quad,
\vspace{0.2cm}
\\
&=
\sum_p
\vert
\braket{\Psi_0\vert \Psi_p}
\vert^2\,
\delta(E-E_p))
\quad.
\end{align}
Here, $C(t)=\braket{\Psi(t/2)\vert\Psi(t/2)}$ is the autocorrelation function and, $t_f$, is the final propagation time, which we here chose as, $t_f=3000\,\mathrm{fs}$. Further, we have vibro-polaritonic eigenergies, $E_p$, and corresponding eigenstates, $\ket{\Psi_p}$, satisfying the time-independent Schr\"odinger equation 
\begin{equation}
\biggl(
\hat{H}_S
+
\hat{H}_C
+
\hat{H}_{SC}
+
\hat{H}_{DSE}
\biggr)
\ket{\Psi_p}
=
E_p
\ket{\Psi_p}
\quad.
\end{equation}
which we solve for the ground state by imaginary time evolution employing the MCTDH method. We provide an illustrative discussion for the molecular dimer, $N=2$, interacting with the cavity mode.
\begin{figure}[hbt!]
\includegraphics[scale=1.0]{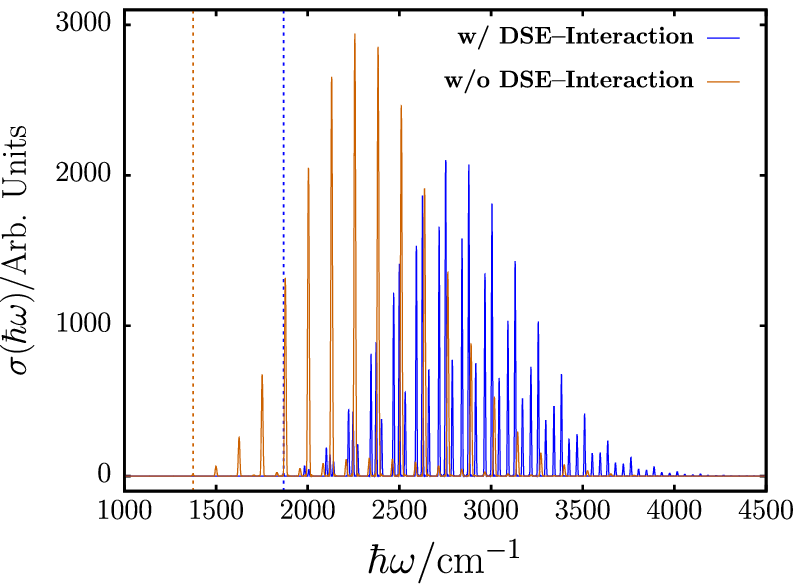}
\renewcommand{\baselinestretch}{1.}
\caption{Vibro-polaritonic density of states, $\sigma(\hbar\omega)$, for $N=2$ hydrogen transfer systems strongly coupled to a single cavity mode as obtained for initial states propagated under the action of the effective Pauli-Fierz Hamiltonian with (blue) and without (orange) DSE-induced dipole-dipole interaction. Vibro-polaritonic ground state energies of $1867\,\mathrm{cm}^{-1}$ (blue) and $1374\,\mathrm{cm}^{-1}$ (orange) are indicated by dashed vertical lines.}
\label{fig.spectra_ensemble}
\end{figure}

In Fig.\ref{fig.spectra_ensemble}, we show $\sigma(\hbar\omega)$ for two different scenarios, where we either account for the full DSE contribution (\textit{cf.} Eq.(6) main text) including the dipole-dipole interaction (blue graph) or where we exclude the dipole-dipole interaction and keep only the diagonal DSE contributions (orange graph). We first realize that in both cases, the envelope approximately reflects a Poisson distribution, which by recalling the cPES shift along the cavity displacement coordinate axis can be rationalized as follows: Instead of considering a shift of the cPES along the negative $x_c$-axis relative to the initial state's location, we can understand the initial state to be shifted relative to the cPES in positive $x_c$-direction instead. A shift of the cavity ground state along $x_c$ results in a coherent state, whose components in terms of cavity number states follow a Poisson distribution\cite{milonni2019}
\begin{align}
P(n_c)
&=
e^{-\braket{\hat{n}_c}}\,
\dfrac{\braket{\hat{n}_c}^{n_c}}{n_c!}
\quad,
\end{align}
with photon number, $n_c$, and length-gauge photon number expectation value, $\braket{\hat{n}_c}$ (\textit{cf.} section \ref{sec.poton_number_operator_length} below). The maximum of $P(n_c)$ is given by, $\braket{\hat{n}_c}$, which we found numerically for the fully interacting situation to be initially, $\braket{\hat{n}_c}_0=7.7$ (\textit{cf.} Fig.\ref{fig.photonnr_unnormalized} for $N=2$). The corresponding density of states is given in Fig.\ref{fig.spectra_ensemble} in blue, with the peak of maximal intensity located at $2752\,\mathrm{cm}^{-1}$ and the first peak corresponding to the vibro-polaritonic ground state located at $1867\,\mathrm{cm}^{-1}$ (indicated by a vertical, blue dashed line). The latter is solely determined by the cavity ground state with $n_c=0$ photons. Thus, we find for the maximum intensity peak and a cavity mode energy of $\hbar\omega_c=126.5\,\mathrm{cm}^{-1}$ approximately $n_c=7$ photons, which agrees well with the numerically obtained $\braket{\hat{n}_c}_0=7.7$. Deviations can be traced back to the strong coupling of cavity mode and molecular systems, such that the initial state is actually not a bare coherent state in the cavity subspace but a correlated light-matter hybrid state with coherent character in the cavity mode.
 
We eventually address differences between blue and orange graphs in Fig.\ref{fig.spectra_ensemble} due to DSE-induced dipole-dipole interactions. We observe the approximate result in orange to be red-shifted relative to the fully interacting result in blue, with ground state energy indicated by a vertical, orange dashed line. We attribute this negative energy shift to an artificial effect induced by the light-matter interaction contribution in the truncated Hamiltonian, which is compensated for when fully accounting for the DSE term\cite{fischer2021}. Note, global energy shifts are not accessible in experiment, where only relative energies are measured and the effects of a global shift cancel. Further, when the full DSE term is included, cavity-induced dipole-dipole interactions between hydrogen transfer systems lead to the formation of additional vibro-polaritonic states. This can be interpreted as a cavity-induced \textit{pseudo} Kasha effect, which emerges from light-matter interaction involving \textit{transverse} cavity fields. We shall point out, that the molecular Kasha effect results from molecular dipole-dipole interactions based on \textit{longitudinal} electric field components (Coulomb-type interaction)\cite{Hestand2018}.

\section{Photon number operator in length-gauge representation}
\label{sec.poton_number_operator_length}
\begin{figure*}[hbt!]
\includegraphics[scale=1.0]{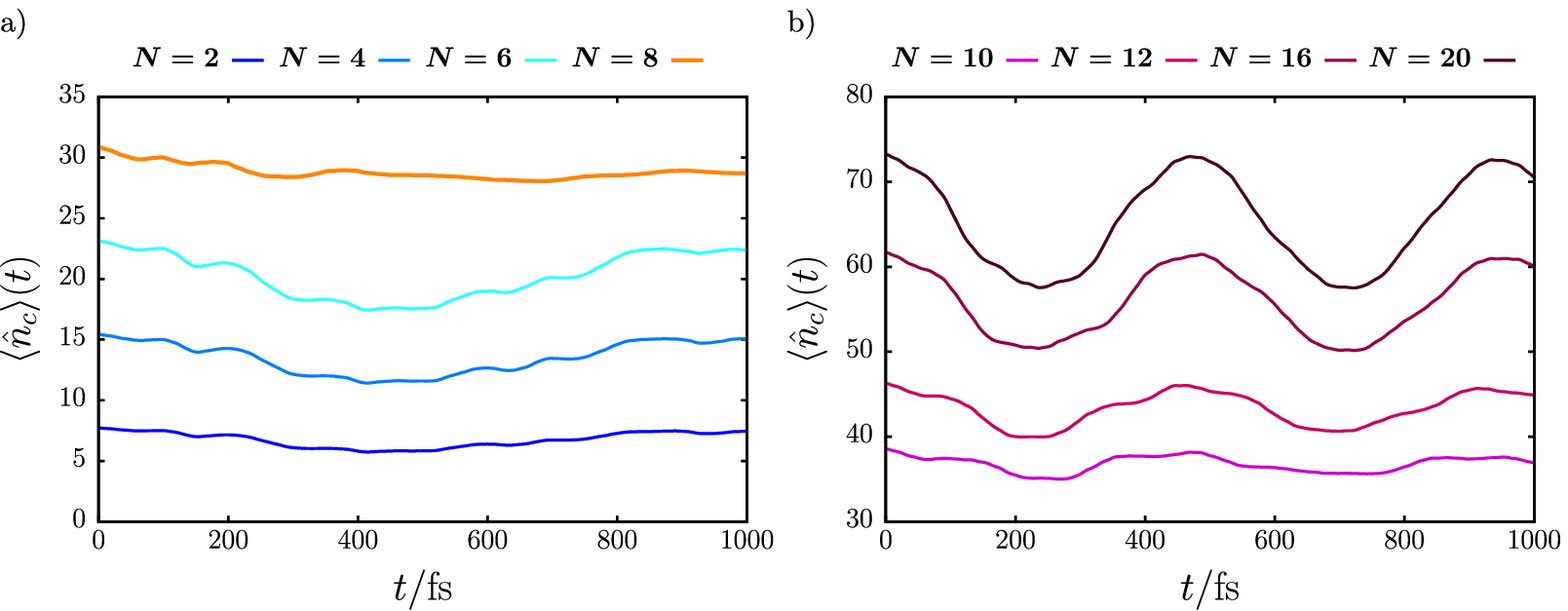}
\renewcommand{\baselinestretch}{1.}
\caption{Time-evolution of photon number expectation value, $\braket{\hat{n}_c}(t)$, as function of ensemble size $N$ for light-matter interaction strength, $\eta=0.05$.} 
\label{fig.photonnr_unnormalized}
\end{figure*}

\setcounter{equation}{0}
In the main text, the expectation value of the photon number operators was used to analyze the cavity-induced H-transfer dynamics. The single-mode photon number operator, $\hat{n}_c=\hat{a}^\dagger_c\hat{a}_c$, can be written in terms of the single-mode cavity Hamiltonian, $\hat{H}_C=\hbar\omega_c\left(\hat{a}^\dagger_c\hat{a}_c+\frac{1}{2}\right)$, as
\begin{equation}
\hat{n}_c
=
\hat{a}^\dagger_c
\hat{a}_c
=
\dfrac{1}{\hbar\omega_c}
\hat{H}_C
-
\dfrac{1}{2}
\quad,
\end{equation}
where, $\hat{a}^\dagger_c$ and $\hat{a}_c$ are bosonic photon creation and annihilation operators, respectively. In length gauge representation, $\hat{n}_c$, takes the form\cite{fischer2021,rokaj2018,schaefer2020,mandal2020}
\begin{align}
\mathcal{S}^\dagger\,
\mathcal{U}^\dagger\,
\hat{a}^\dagger_c
\hat{a}_c\,
\mathcal{U}\,
\mathcal{S}
&=
\dfrac{1}{\hbar\omega_c}
\left(
\mathcal{S}^\dagger\,
\mathcal{U}^\dagger\,
\hat{H}_C\,
\mathcal{U}\,
\mathcal{S}
\right)
-
\dfrac{1}{2}
\quad, 
\end{align}
with unitary operator, $\mathcal{U}=\exp\left[\frac{\mathrm{i}}{\hbar}\,\hat{A}\,d(q)\right]$, mediating the Power-Zienau-Woolley (PZW) transformation as generated by the molecular dipole moment, $d(q)$, and the transverse (single-mode) vector potential, $\hat{A}=\frac{g}{\omega_c}\left(\hat{a}^\dagger_c+\hat{a}_c\right)$. A second unitary rotation mediated by, $\mathcal{S}=\exp\left[\mathrm{i}\frac{\pi}{2}\,\hat{a}^\dagger_c\hat{a}_c\right]$, acts exclusively on the cavity mode subspace and provides a real light-matter interaction term, $\hat{H}_{SC}$. Under $\mathcal{U}$ and $\mathcal{S}$, photon creation and annihilation operators transform as
\begin{align}
\mathcal{S}^\dagger\,
\mathcal{U}^\dagger\,
\hat{a}^\dagger_c\,
\mathcal{U}\,
\mathcal{S}
&=
-
\mathrm{i}\,
\hat{a}^\dagger_c
-
\dfrac{\mathrm{i}}{\hbar}
\dfrac{g}{\omega_c}
d(q)
\quad,
\vspace{0.2cm}
\\
\mathcal{S}^\dagger\,
\mathcal{U}^\dagger\,
\hat{a}\,
\mathcal{U}\,
\mathcal{S}
&=
+
\mathrm{i}\,
\hat{a}_c
+
\dfrac{\mathrm{i}}{\hbar}
\dfrac{g}{\omega_c}
d(q)
\quad.
\end{align}
Employing the latter identities, the transformed number operator turns with
\begin{widetext}
\begin{align}
\hbar\omega_c\,
\left(
\mathcal{S}^\dagger\,
\mathcal{U}^\dagger\,
\hat{a}^\dagger_c
\hat{a}_c\,
\mathcal{U}\,
\mathcal{S}
\right)
&=
\hbar\omega_c
\left(
-
\mathrm{i}\,
\hat{a}^\dagger_c
-
\dfrac{\mathrm{i}}{\hbar}
\dfrac{g}{\omega_c}
d(q)
\right)
\left(
+
\mathrm{i}\,
\hat{a}_c
+
\dfrac{\mathrm{i}}{\hbar}
\dfrac{g}{\omega_c}
d(q)
\right)
\quad,
\vspace{0.2cm}
\\
&=
\hbar\omega_c
\left(
\hat{a}^\dagger_c
\hat{a}_c
+
\dfrac{g\,d(q)}{\hbar\omega_c}
\left(
\hat{a}^\dagger_c
+
\hat{a}_c
\right)
+
\dfrac{g^2}{(\hbar\omega_c)^2}
\,d^2(q)
\right)
\quad,
\vspace{0.2cm}
\\
&=
\hbar\omega_c\,
\hat{a}^\dagger_c
\hat{a}_c
+
g\,d(q)
\left(
\hat{a}^\dagger_c
+
\hat{a}_c
\right)
+
\dfrac{g^2}{\hbar\omega_c}
\,d^2(q)
\end{align}
\end{widetext}
as well as identities, $x_c=\sqrt{\frac{\hbar}{2\omega_c}}\left(\hat{a}^\dagger_c+\hat{a}_c\right)$, and, $\hat{p}_c=\mathrm{i}\sqrt{\frac{\hbar\omega_c}{2}}\left(\hat{a}^\dagger_c-\hat{a}_c\right)$, into
\begin{widetext}
\begin{align}
\mathcal{S}^\dagger\,
\mathcal{U}^\dagger\,
\hat{a}^\dagger_c
\hat{a}_c\,
\mathcal{U}\,
\mathcal{S}
&=
\dfrac{1}{\hbar\omega_c}
\left(
\underbrace{
\dfrac{\hat{p}^2_c}{2}
+
\dfrac{\omega^2_c}{2}\,x^2_c}_{=\hat{H}_C}
+
\underbrace{\sqrt{\dfrac{2\omega_c}{\hbar}}\,g\,d(q)\,x_c}_{=\hat{H}_{SC}}
+
\underbrace{\dfrac{g^2}{\hbar\omega_c}\,d^2(q)}_{=\hat{H}_{DSE}}
\right)
-
\dfrac{1}{2}
\quad,
\vspace{0.2cm}
\\
&=
\dfrac{1}{\hbar\omega_c}
\left(
\hat{H}_C
+
\hat{H}_{SC}
+
\hat{H}_{DSE}
\right)
-
\dfrac{1}{2}
\quad.
\end{align}
\end{widetext}
This latter expression enters the cavity photon number expectation value, $\braket{\hat{n}_c}$, in Eq.(14) of the main text. The same arguments generalize to $\hat{n}_c$ for ensembles of $N$ molecules as the unitary operator mediating the PZW transformation, $\mathcal{U}_N=\prod^N_i\,\mathcal{U}_i$, factorizes due the form of the ensemble dipole function, $d(\underline{q})=\sum^N_i\,d(q_i)$. In Fig.\ref{fig.photonnr_unnormalized}, we eventually provide the time-evolution of the bare photon number operator expectation value without normalization to the initial value.
